\documentclass[journal,comsoc]{IEEEtran}

\usepackage[T1]{fontenc}
\usepackage{graphicx}
\usepackage{diagbox}
\usepackage{textcomp}
\usepackage{xcolor}
\usepackage{caption}
\usepackage{subcaption}
\usepackage{stfloats}

\usepackage{framed}
\usepackage{tcolorbox}
\usepackage{adjustbox}
\usepackage[figurename=Figure, labelfont= bf]{caption}
\graphicspath{{../figures/}}

\ifCLASSINFOpdf
\else
\fi

\usepackage{amsmath}
\usepackage[cmintegrals]{newtxmath}
\makeatletter
\def\@citex[#1]#2{\leavevmode
\let\@citea\@empty
\@cite{\@for\@citeb:=#2\do
{\@citea\def\@citea{,\penalty\@m\ }%
\edef\@citeb{\expandafter\@firstofone\@citeb\@empty}%
\if@filesw\immediate\write\@auxout{\string\citation{\@citeb}}\fi
\@ifundefined{b@\@citeb}{\hbox{\reset@font\bfseries ?}%
\G@refundefinedtrue
\@latex@warning
{Citation `\@citeb' on page \thepage \space undefined}}%
{\@cite@ofmt{\csname b@\@citeb\endcsname}}}}{#1}}
\makeatother

\begin{document}

\title{\sc{Satellite Based Computing Networks with Federated Learning}}

\author{Hao~Chen,~
	Ming~Xiao,~
    and~Zhibo~Pang
	\thanks{This work was supported by ERA-NET Smart Energy Systems SG+ 2017 Program, "SMART-MLA" with Project number 89029 (and SWEA number 42811-2), Swedish Research Council Project entitled "Coding for Large-scale Distributed Machine Learning", Swedish Foundation for International Cooperation in Research and Higher Education (STINT), project "Efficient and Secure Distributed Machine Learning with Gradient Descend",  and FORMAS project entitled "Intelligent Energy Management in Smart Community with Distributed Machine Learning", number 2021-00306. Zhibo Pang’s work is partly funded by the Swedish Foundation for Strategic Research (SSF) through the project APR20-0023. (Corresponding author: Ming Xiao.)}
    
    \thanks{Hao Chen, Ming Xiao and Zhibo Pang are with the School of Electrical Engineering and Computer Science, Royal Institute of Technology (KTH), 10044 Stockholm, Sweden (email: \{haoch, mingx, zhibo\}@kth.se).
	
	Zhibo Pang is also with Department of Automation Technology, ABB Corporate Research Sweden, Vasteras, Sweden (email: pang.zhibo@se.abb.com).	
 	}
    }

\maketitle

\section*{\textbf{Abstract}}
Driven by the ever-increasing penetration and proliferation of data-driven applications, a new generation of wireless communication, the sixth-generation (6G) mobile system enhanced by artificial intelligence (AI), has attracted substantial research interests.
Among various candidate technologies of 6G, low earth orbit (LEO) satellites have appealing characteristics of ubiquitous wireless access.
However, the costs of satellite communication (SatCom) are still high, relative to counterparts of ground mobile networks.
To support massively interconnected devices with intelligent adaptive learning and reduce expensive traffic in SatCom, 
we propose federated learning (FL) in LEO-based satellite communication networks.
We first review the state-of-the-art LEO-based SatCom and related machine learning (ML) techniques, and then analyze four possible ways of combining ML with satellite networks. 
The learning performance of the proposed strategies is evaluated by simulation and results reveal that FL-based computing networks improve the performance of communication overheads and latency.
Finally, we discuss future research topics along this research direction.

\IEEEpeerreviewmaketitle

\section*{\textbf{Introduction}}
With emerging smart applications (e.g., augmented reality/virtual reality (AR/VR) and digital twin), an unprecedented amount of data are generated through the connected Internet of things (IoT) devices, ranging from mobile sensors, wearable devices, smartphones to connected vehicles and unmanned serial vehicles (UAVs).
It is reported in \cite{iot_device} that the active number of IoT devices is expected to be over 75 billions by 2025.
Despite of the wide deployment of mobile networks, many devices in remote areas (e.g., deserts or ocean), still do not have connection services.
It is shown that terrestrial wireless networks only cover about 20 percent of the global land area and less than 6 percent of the surface of the earth due to extreme terrain or communication distance and commercial and engineering difficulties \cite{cover_tes}. 
To achieve the global coverage and seamless connectivity to support various computing tasks such as artificial intelligence (AI), there is consensus in academy and industries that satellite communication (SatCom) networks are viable complementary alternative to terrestrial networks.
SatCom can provide wireless access services with the benefits of high throughputs and global broadband coverage to a diverse user population, enhancing the network performance and offloading the terrestrial traffic efficiently as well as real-time services, especially in distant and unserved areas {\color{black}\cite{6g, sat_com1}}.

Compared to traditional geostationary earth orbit (GEO) satellites in a altitude of 36, 000 km above the earth, the constellation of networks consisting of low earth obit (LEO) satellites, can provide broadband network services and has attracted a lot of interests in both industries and academy recently \cite{sat_IoT}.
Typically, LEO satellite constellation is deployed at a height of 500-2000 km from Earth, which could offer faster communications (i.e., lower propagation latency), lower energy consumption, suppression of signalling attenuation by allowing more tightly focused beams to be projected on the ground.
Ongoing LEO constellation projects, including SpaceX Starlink, OneWeb, Telesat Lightspeed, LeoSat, and Amazon Project Kuiper, will launch over 46, 100 satellites over Earth in near future, aiming to deploy an ultra-dense constellation and jointly with the traditional land mobile operators to support seamless and high-capacity communication services \cite{more_sat}.
However, the integration of terrestrial and satellite networks, poses new challenges for the whole protocol stack, and many topics are still open for research \cite{fang20215g, sate_edge}.

In the past decades, we have witnessed the revolutionary breakthroughs brought by AI, which rely heavily on large quantity and high quality data. 
The long-lasting data originated from massive IoT devices are commonly produced and stored in a distributed manner for many applications (e.g., smart grids, environment monitoring).
It is often impractical or inefficient to collect and send all data to a centralized location due to the limitations of communication resources or latency.
Meanwhile, with increasing awareness of data privacy and security (e.g., GDPR in Europe \cite{gdpr} to protect sensitive information such as personal medical or financial records, and confidential business notes), data processing and inference close to the the sources or devices have become a pivotal role in avoiding high latency, communication overheads and information leakage. 
Contrary to traditional machine learning (ML) with raw data transferred and processed in a central entity (e.g., the cloud), decentralized ML by leveraging edge computing and intelligence with distributed data kept locally is one alternative solution to data analysis especially for large-scale ML models.
As one of important distributed ML schemes, federated learning (FL), was first coined by Google in 2016 \cite{fedavg}, which pushes the computation of AI applications into lots of end devices without violating the privacy.
The main benefits of FL include communication efficiency and data privacy by only transmitting model parameters instead of raw data. 
This makes it very suitable for the future sixth-generation (6G) system.
In fact, FL-based approaches have already been deployed by major service providers, and are making a significant impact on supporting communication-efficient and privacy-sensitive applications where the training data are distributed at the edge. 
Statistical models of applications such as next-word prediction, face detection, and voice recognition, can be inferred via jointly learning user behavior across a large pool of mobile phones without sending the raw data to a central server.

Motivated by the above observations, we propose FL-based computing networks in satellite constellation.
The remainder of the article is organized as follows.
We start by giving preliminary knowledge on satellite communication networks, classical centralized learning and federated learning. 
Then the pros and cons of the four methods for integrating learning with satellite networks are analyzed.
We present simulation results to validate the efficiency of proposed architectures. 
Finally, we conclude our work and suggest potential research directions.

\section*{\textbf{Overview}}
\subsection*{\centering\sc{Low Earth Orbit Satellite Communication}}
SatCom, in particular LEO satellite constellation, is a promising technique to satisfy the 6G mobile requirements in terms of global coverage and ubiquitous connectivity in the near future, particularly for under-served areas like isolated areas, where constantly lack of communication infrastructure. 
It is expected that SatCom will be an integral part in the Future Communications Infrastructure (FCI).
\begin{figure} [t] 
	\vskip 0.2in
	\begin{center}
		\centerline{\includegraphics[width=88mm]{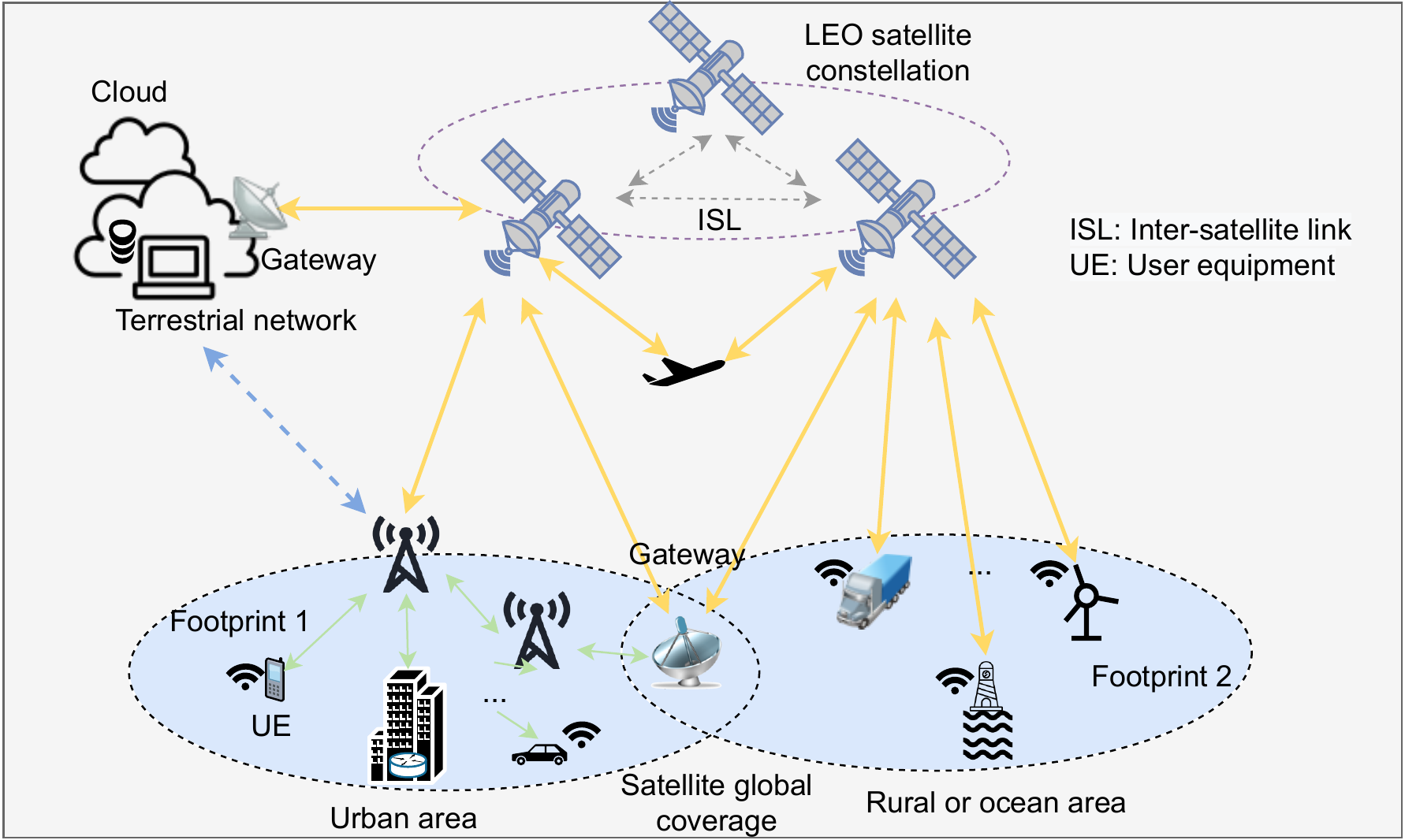}}
		\caption{The architecture of STN in crowed and unserved areas.}
	    \label{fig:sat_archi}
	\end{center}
	\vskip -0.2in
\end{figure} 

\textbf{LEO-based SatCom:} Figure 1 shows the framework of LEO based satellite-terrestrial networks (STN) with user equipment (UE) in crowded and unserved areas.
UEs access the STN from different service areas (i.e., footprints), each of which can be seen as a circular area on the earth surface.
In general, LEO-based SatCom consists of LEO satellite constellation in the space segment, and multiple gateways and terrestrial backbone networks in the ground segment.
Specifically, terrestrial backbone networks connect to external networks (e.g., Internet or cloud), which are responsible for various tasks such as mission planning, network control, resource management, cloud computing and centralized learning, to name a few.
Gateways are mainly used to provide connection for end-user devices (i.e., UEs) and perform tracking and upload operations.
To provide seamless global coverage and connectivity, LEO satellites have inter- and intra- plane connections via inter-satellite links (ISL) and serve as the relays for data transmission. 
In addition, a part of satellites are equipped with on-board processing and storage capabilities for satellite-borne computing.

\textbf{Satellite spectrum:}
SatCom can operate in the super high frequency (SHF) band, in particular between 1-50 GHz.
Then different frequency bands are suitable for various climate conditions, UEs and services.
Legacy frequency bands below 6 GHz, that is, L-band (1-2 GHz), S-band (2-4 GHz), and C-band (4-6 GHz), have been predominately operated for applications such as satellite TV, GPS services and space missions \cite{freq_band1}. 
However, they cannot satisfy the unprecedentedly stringent capacity demands envisioned by the future 6G wireless services.
Millimeter-wave (mmWave), such as Ku-band (12-18 GHz) and Ka-band (26.5-40 GHz), have thus been recently adopted in satellite systems \cite{freq_band2}.

\textbf{SatCom use cases:}
For past decades, terrestrial networks have demonstrated their superior capabilities of supporting mobile device connectivity in presence of high user density.
However, existing terrestrial cellular technologies (e.g., LTE, 4G, and 5G), may not meet higher requirements of future wireless applications, such as resilience, security, and availability, and are vulnerable to natural disasters and terrorist attacks \cite{6g}.
On the other hand, SatCom has attracted substantial research interests in multiple critical vertical applications.
For instance, in the agriculture sector, to increase yields and productivity of farm, several companies, such as Milk Smarts, are integrating satellite backhauling of low-power wide-area (LPWA) to new solutions for carrying out precision farming along with advancing its storage and distribution.
In healthcare sector, telemedicine or remote health consulting to improve the quality of people life is gaining prominence across the world, especially in Europe.
To combine the remote network together into an integrated medical center, LEO satellites can be fully used such that healthcare services in remote locations such as testing and diagnosis in remote clinics can be significantly expanded.
In public protection and disaster relief sector, SatCom provides the full support of guaranteed communication under the circumstance that terrestrial networks function abnormally (e.g., in natural disasters such as a flooding or an earthquake).
Besides, SatCom has a crucial role in forest fire monitoring since forest fire usually happens in remote or mountain areas where terrestrial networks can barely be deployed.

\subsection*{\centering\sc{Machine Learning and Federated Learning}}
Machine learning for massively connected IoTs has traditionally been focused on centralized learning (CL), where a powerful artificial neural network (ANN) model is often trained by uploading all raw data from each connected UE to the cloud and this generic model can then been distributed and applied to all UEs. 
Amazon Web Services, Google Cloud, and Microsoft Azure are the typical ML as-a-service providers, where models can be deployed and used at a large scale.
However, these learning schemes face the challenges of high communication costs and security.
To address the problems, federated learning (FL), has been proposed as a communication-efficient and privacy-preserving paradigm of decentralized learning, where participated UEs can collaboratively build a shared learning model while leaving the training data locally \cite{fedavg}. 
In particular, a UE computes its update to the current global model on its local training data and periodically transmits model parameters to a central server, where the local models are then aggregated into global models, which are sent back according to aggregation strategies such as the federated averaging algorithm (FedAvg) \cite{fedavg}.
This process is repeated until a target accuracy of the learning model is reached. 
In such a way, the user data privacy is well protected since local training data are not shared, and the communication loads are relieved since only model parameters are exchanged instead of a large amount of raw-data samples, which differs FL from conventional CL in data acquisition, storage, training, and inference.
\section*{\textbf{Federated Learning in Satellite Networks}}

In this section, we investigate the possible ways to implement FL in LEO-based satellite constellation networks and then we summarize benefits of FL in SatCom.
\begin{figure*} [t] 
	\vskip 0.2in
	\begin{center}
		\centerline{\includegraphics[width=160mm]{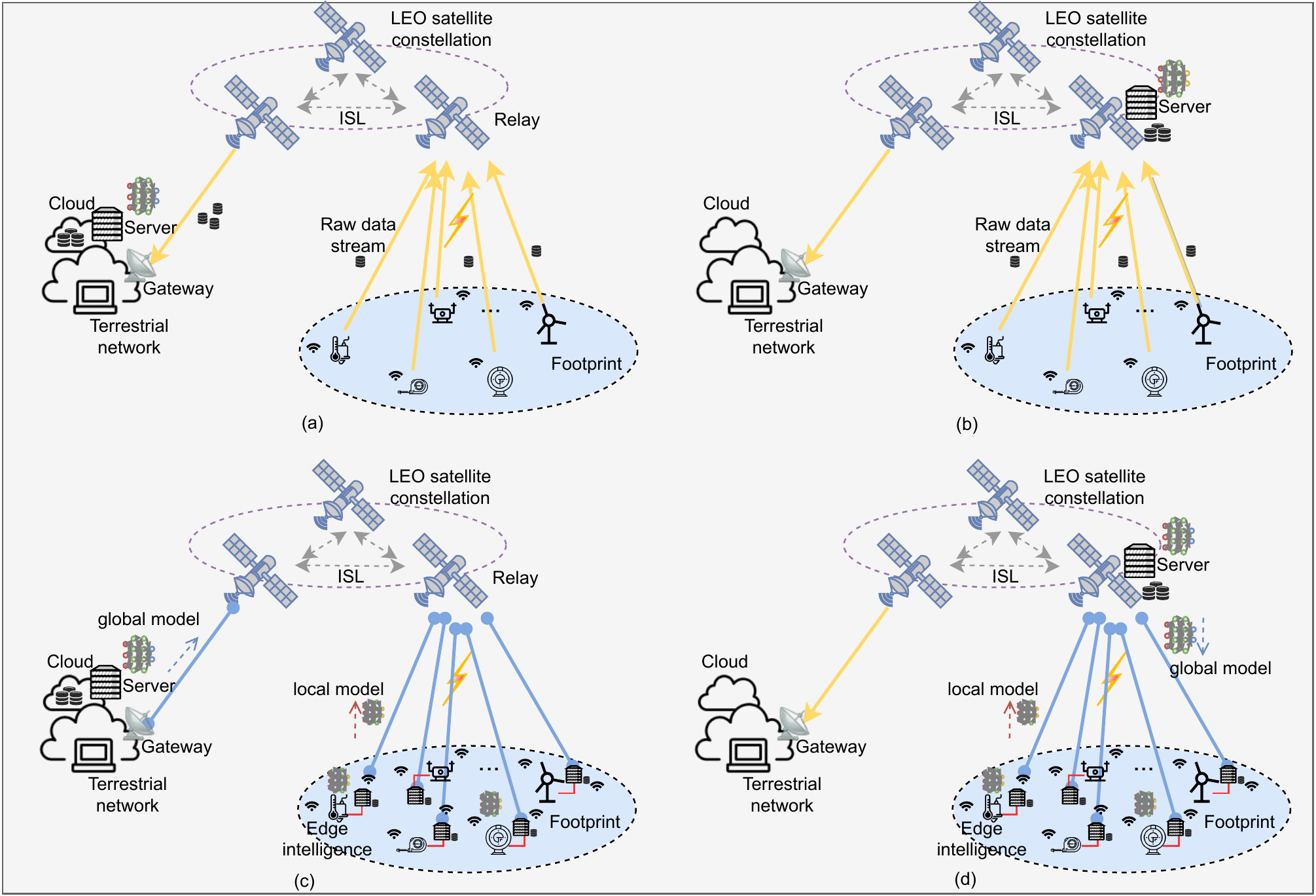}}
		\caption{ML in LEO-based SatCom networks: (a) CL with satellites as relays; (b) CL with computing servers allocated at satellites; (c) FL with satellites as relays; (d) FL with satellites as the servers.}
	    \label{fig: stn_archi}
	\end{center}
	\vskip -0.2in
\end{figure*} 
\subsection*{\centering\sc{Where to learn?}}
\begin{table*}[!h]
    \label{tab:comparison}
    \centering
    \fontsize{9}{8}\selectfont
    \begin{tabular}{|c|c|c|c|c|c|}
        \hline 
        \diagbox{Property}{Architecture}
        & Remote cloud learning & On-board satellite learning & Federated learning 1 & Federated learning 2 \\
        \hline
        
        \hline
        Communication overheads &   Very high  & High & Moderate & Low \\ 

        \hline 
        Deployment costs & Low & Moderate & Moderate & High\\

        \hline
        Privacy and security & Very low & Low & High & High \\

        \hline
        Latency & Very high & High & Moderate & Low\\ 
                            
        \hline
        Energy consumption & High & Very high & Low & Moderate\\ 

        \hline
        Feasibility & High & Moderate & Very high & High\\
        \hline
        
    \end{tabular}
    \caption{Comparison between the four potential modes{\color{blue}.}}
\end{table*}

In LEO-based STN, UEs may jointly learn a data-driven model,
among which UEs are served by terrestrial base stations (TBSs) when they are within the coverage of TBSs and can gain access via satellites when TBSs are not available and work as relays to forward signals between UEs and LEO satellites.
In what follows, to simplify illustration, we specifically consider a SatCom that consists of LEO satellite constellation, a cloud, a server, and multiple UEs. 
Provided that learning in such a wide area is established with the local data information of UEs. 
As depicted in Fig. 2, the four possible learning strategies in SatCom are elaborated as below.
For illustration, properties of four possibilities of integrating ML in SatCom are summarised in Table I.

\textbf{Mode 1 (Remote cloud learning):} In Fig. 2a, the central server is deployed in cloud and raw-data stream of UEs are then directly transmitted to the cloud server.
Thus, a global model is developed by classical CL in the cloud with the centralized processing and inference and then is applied to UEs.
In this scenario, LEO satellites are utilized as relays to re-transmit the traversed data stream.
Mode 1 is straightforward to integrate into the current communication system due to the conceptualization and development of hybrid STN since 1964 [7].
However, it inevitably incurs a longer latency, which is not favorable scheme for real-time applications since lower end-to-end latency is one of the main technical concerns.

\textbf{Mode 2 (On-board satellite learning)}: Different from Mode 1, the computing servers are deployed in satellites with close proximity to UEs instead of remote cloud, as displayed in Fig. 2b.
Compared to Mode 1, the latency of this mode by propagation delay and transmission time, is reduced because communication between the cloud and satellites is avoided.
Meanwhile, the probability of information leakage is decreased due to the fewer number of communication hops.
Thus, Mode 2 is quite suitable for delay-sensitive applications such as military communications.
However, this scheme requires that satellites are equipped with extensive computation and storage hardware.
By embedding a dedicated server in the on-board satellite, one main potential drawback of this mode is economically expensive.
Most importantly, intensive on-board computation and training consume lots of precious energy of satellites, which may not be practical when the energy supply of satellites is strictly limited.

\textbf{Mode 3 (Federated learning 1)}: As shown in Fig. 2c, 
SatCom builds a generic model via FL without raw-data sharing. 
Although the parameter server is still allocated in the remote cloud as Mode 1, the protection of data privacy and security of UEs can be significantly enhanced by Mode 3.
Since only model parameters are transmitted among UEs, satellites and cloud and the sheer bulk of raw data are not exchanged in the networks. 
Thus, Mode 3 dramatically improves latency and communication overheads.
Further, it is one of most flexible and robust modes since UEs have more flexibility to decide on participating for building the global model whenever a straggler event happens due to the disconnection of UEs or poor wireless connection.
However, this mode leads to relatively higher deployment costs in UEs than CL-based approaches, since the capabilities of local computing and training are prerequisites and learning tasks are distributed to UEs. 
The characteristic of iterative process of FL may also increase the cumulative communication overhead although in most cases it is still much lower than CL-based strategies.
Overall, Mode 3 is quite suitable for practical implementation and maintenance.

\textbf{Mode 4 (Federated learning 2)}: When it is feasible to deploy parameter servers in on-board satellites and run FL without data sharing, we have Mode 4, as illustrated in Fig. 2d.
As we can see from Table I, communication overheads, information leakage and latency are the lowest among all strategies because the parameter sever is not only closer to UEs, but the number of visited intermediate network nodes (i.e., satellite, gateway, etc.) is negligible.
It is also insignificant from the current architecture on the burden of the remote cloud.
Note that on-board satellite aggregation in Mode 4 evokes more energy consumption compared to Mode 3, but is more energy efficient than Mode 2.

\subsection*{\centering\sc{What can FL bring in LEO-based SatCom?}}
In LEO satellite communication networks, UEs from multiple areas (e.g., forests and urban woodlands) may collaboratively build a shared learning model (fire monitor) via the FL framework. 
Similar to the current operating application of Gboard on Android \cite{gboard}, FL in SatCom may have the following advantages.
Firstly, FL-based SatCom has fast response. 
Data processing and learning are at the terminal of the SatCom and closer to the UEs, thereby reducing the response time of decision-making. 
Secondly, compared to classical learning approach, the parameter server in FL is flexible enough to decide whether to accept a burst of UEs to accelerate ML process or directly ignore UEs with poor or limited connectivity condition. 
Meanwhile, UEs in SatCom have the high level of flexibility to participate or leave the local processing and cooperation among them.
Thirdly, data privacy and security of individual entities in SatCom are enhanced. 
Training and predicting can be completed without sharing data.
Thus, sensitive data of UEs will not be exposed or exported to servers or even third parties.
Finally, communication efficiency in FL-based SatCom is improved compared with those of CL-based architectures.

\section*{\textbf{Performance Evaluation}}
\begin{figure*}[t] 
\begin{framed}
 	\centering
 	\vskip -0.1in
	\subfloat[]{\hspace*{1mm}\includegraphics[width=61mm]{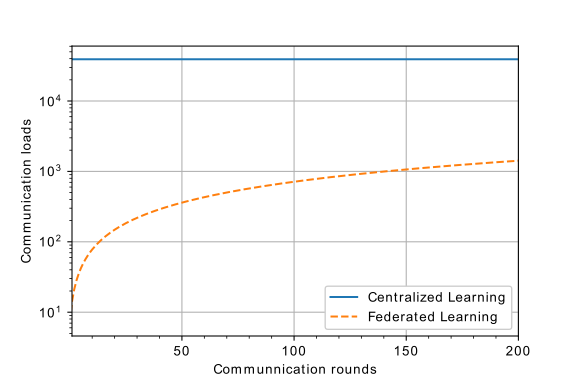}\label{fig6:result1}}
	\subfloat[]{\hspace*{-2mm}\includegraphics[width=61mm]{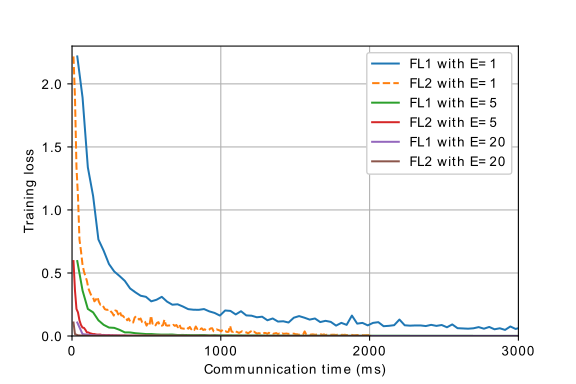}\label{fig6:result2}}
	\subfloat[]{\hspace*{-2mm}\includegraphics[width=61mm]{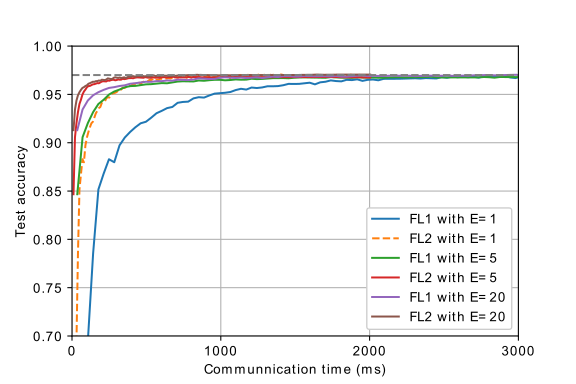}\label{fig6:result3}
	}
	\end{framed}
    \caption{Performance of FL in LEO-based satellite communication networks on the MNIST dataset: a): communication overhead comparison between CL and FL; b) training loss vs. communication time for many local epochs (large $E$); c) test accuracy vs. communication time for many local epochs (large $E$).}
    \label{fig_performance_mnist}
 \end{figure*}

To assess the performance of the proposed distributed learning strategies in satellite communication networks, we consider a SatCom similar to the example network shown in Fig. \ref{fig: stn_archi}.
A typical Iridium system consisting of 66 LEO satellites is utilized to emulate the LEO constellation to provide cellular-like service in areas
where terrestrial cellular service is unavailable.
Totally $N=100$ UEs identically equipped with computing and memory are assumed to be dispersed on the under-served areas for training a model of interest with the following parameters: the propagation delay between directly-connected LEO satellite and UE/cloud is assumed to be 5 ms, and the considered ISL delay follows the uniform distribution and is in the range of $[5, 15]$ ms, the data rate for upstream and downstream in SatCom could be up to 8 Mbps \cite{delay_sat}.
The well-known dataset, MNIST dataset \cite{fedavg}, is used to train FL model for digit recognition tasks.
The dataset consists of 10 classes of $28\times28$ gray-scale images of handwritten digits with 60, 000 training samples and 10, 000 test samples, respectively.
During model training, the training data are partitioned over $N$ blocks, each of which is only viable at one UE that are independently and identically distributed. FedAvg \cite{fedavg} is chosen for the FL implementation.
Thus, the loss function $f(\textbf{w}^t)$ of the global model at the $t{\text{-th}}$ communication round can be calculated by
\begin{equation}
    f(\textbf{w}^t) = \sum _{i=1}^N \frac{n_i}{n} F_i(\textbf{w}_i^t),
\end{equation}
where $N$ is the total number of involved UEs in one FL task, $n$ is the total number of data samples for all involved UEs and $\textbf{w}^t$ represents the global FL model parameter, while $n_i$, $\textbf{w}_i^t$, and $F_i(\textbf{w}_i^{t})$ are the number of data samples, the local FL model parameter and the loss function for the $i{\text{-th}}$ UE, respectively.
Considering the slow responding due to the opportunistic offline or slow or expensive connections, only ten percent of UEs are randomly selected to join in the learning process in each learning round.
For the target model, we consider a simple convolutional neural network (CNN) with two $5\times5$ convolutional layers (the first with 10 channels, the second with 20, and each followed by ReLU function), a fully connected layer with 320 units and ReLU activation, and a final softmax output layer (8, 480 total parameters).
For fair comparison, the hyperparameters for learning approaches are tuned and kept the same in different experiments.
Our model was simulated by PyTorch in Python and experiments were carried out on an Intel CPU @2.3 GHz (16-GB RAM) laptop.

We first plot in Fig. 3a the communication loads of FL-based architectures (i.e., Mode 3 and Mode 4) vs. CL-based architectures (i.e., Mode 1 and Mode 2) in terms of communication rounds. 
Communication rounds are defined as the cumulative upload number of communication.
Communication loads are measured by the number of transmitted packets during model training. 
For CL-based approach, the communication load is equal to the total number of packets for transmitting raw-data stream, whilst it is the accumulative number of packets for transmitting model parameters in FL-based approach, where it is proportional to the number of participated UEs and the number of communication round.
It is observed that communication loads of CL-based approaches keep constant as communication rounds increase and they increase gradually in cases of Mode 3 and Mode 4.
This is due to the fact that in CL-based ways (i.e., Mode 1 and Mode 2), raw data are uploaded into a centralized entity (i.e., either cloud or on-board satellite) before learning, whilst Mode 3 and Mode 4 refer to iterative process with model parameters sharing in each communication round.
It is also noted that the proposed FL-based approaches can significantly reduce the communication loads compared with traditional CL-based approaches in SatCom.

In Fig. 3b and Fig. 3c, we evaluate the learning performance of the proposed FL-based architectures with respect to communication time (latency) as well as number of local epoch. 
We define epoch as the the number of training passes each UE makes over its local dataset on each communication round (denoted as E in figures) and communication time as the sum of transmission time of uploading and downloading model parameters and propagation time.
In fact, the link speed among UE and satellites, is sufficiently high (e.g., 400 Mbps in OneWeb \cite{WinNT}), the transmission time is minor and can be negligible. 
In such cases, the communication time is mainly determined by the propagation time.
For two proposed potential modes in SatCom, training loss and test accuracy are evaluated in Fig. 3b and Fig. 3c, respectively.
As can be seen from Fig. 3b, the convergence speed of the proposed Mode 3 (i.e., FL 1 in figures) is slower than that of Mode 4 (i.e., FL 2 in figures).
This is because that in each communication round, Mode 3 suffers from extra communication costs such as ISL delay and communication time among cloud and LEO satellites compared to Mode 4.
Also, it is demonstrated in Fig. 3b that by increasing $E$ to add more local updating of UE on each communication round, the communication time can be significantly reduced.
In terms of test accuracy, we observe the same learning performance as well.
This proves that the proposed mechanisms can foster the availability and low-cost deployment of FL in LEO satellite constellation and alleviate their long communication time to build a delay-sensitive learning.

\section*{\textbf{Challenges in Future Research}}

For higher connectivity and bandwidth-limited communications, FL-based computing networks in SatCom proposed in this article may face new technical challenges as follows:

\textbf{Privacy and security:}
Although privacy and security have been among the initial objectives of adopting FL in SatCom as a pertinent solution, the distributed characteristic has raised additional issues to be addressed, such as revealing sensitive information via poisoning local data and shared models.
There exist lots of challenges though recent efforts have adopted different privacy-based approaches.
For instance, when differential privacy is introduced, various levels of artificial noise are also injected to improve privacy, which directly degrades the accuracy of the built model. 
Further, if a malicious server exists, the sensitive information of UEs can still be leaked.
As a result, it is urgently demanded to develop a robust privacy-preserved and secure system, where formal privacy and security are guaranteed with limited accuracy loss.

\textbf{Resource management:}
Considering that most LEO satellite resources are under-utilized in current networks due to periodical change of the coverage of the satellite. 
It is thus inevitable to maximize resource utilization particularly under scarce satellite resources with limited spectrum and orbits.
Under such circumstances, software defined network (SDN), network function virtualization (NFV), and ML offer the possibilities to break through the above difficulty.
Dynamic resource management by leveraging an organic combination of SDN and NFV, can be more flexible and reconfigurable in real events.

\textbf{Communication reduction:}

Though current satellites provide much more bandwidth than those of the past few years, their bandwidth still cannot be comparable to the terrestrial media, in particular, the optic-fibers.
Owing to the feature of iterative model updates, communication overhead will be still a major issue under satellite based computing networks with FL.
To achieve pertinent, sustainable, and efficient FL-based solutions, it still needs to pave a way to communication efficiency of FL like promising strategies of communication compression technique via quantization or sparsification.

\section*{\textbf{Conclusions}}
FL over wireless communication networks can dramatically improve learning performance to satisfy ever increasing requirements in data privacy and communication overheads in future 6G networks.
We discussed the role of FL in addressing the challenges in LEO-based SatCom.
The proposed schemes considered the preliminary combination of FL and LEO satellite systems.
Performance evaluation indicates that the proposed integration of FL under LEO satellite constellation has practical accuracy on the MNIST dataset, and communication overheads of FL-based approaches are much smaller than CL-based approaches.
In addition, we have noticed several challenges regarding privacy and security, resource management and communication overheads in future research.

\bibliography{ref}
\bibliographystyle{IEEEtran}

\ifCLASSOPTIONcaptionsoff
  \newpage
\fi

\section*{\textbf{Biographies}}
\begin{IEEEbiographynophoto}{\sc{Hao Chen}} (haoch@kth.se)
received the B.Sc. degree in communication engineering and the M.Sc. degree in electronic and communication engineering from Soochow University, China, in 2014 and 2017, respectively.
He is currently pursuing the Ph.D. degree with the School of Electrical Engineering and Computer Science, KTH Royal Institute of Technology, Stockholm, Sweden.
His current research interests include distributed machine learning, distributed optimization and edge computing.
\end{IEEEbiographynophoto}

\begin{IEEEbiographynophoto}{\sc{Ming Xiao}}[S'02, M'07, SM'12] (mingx@kth.se)
received Bachelor and Master degrees in Engineering from the University of Electronic Science and Technology of China, ChengDu in 1997 and 2002, respectively. He received Ph.D. degree from Chalmers University of Technology, Sweden in November 2007.
From 1997 to 1999, he worked as a network and software engineer in ChinaTelecom. From 2000 to 2002, he also held a position in the SiChuan communications administration. From November 2007 to now, he has been in the department of information science and engineering, school of electrical engineering and computer science, Royal Institute of Technology, Sweden, where he is currently an Associate Professor.
\end{IEEEbiographynophoto}

\begin{IEEEbiographynophoto}{\sc{Zhibo Pang}}[M'13, SM'15] (pang.zhibo@se.abb.com)
received MBA in Innovation and Growth from University of Turku in 2012 and PhD in Electronic and Computer Systems from the Royal Institute of Technology (KTH) in 2013. He is currently a Senior Principal Scientist at ABB Corporate Research Sweden, and Adjunct Professor at the University of Sydney and the Royal Institute of Technology (KTH). He is a Senior Member of IEEE and Co-Chair of the Technical Committee on Industrial Informatics. He is Associate Editor of IEEE Transactions on Industrial Informatics, IEEE Journal of Biomedical and Health Informatics, and IEEE Journal of Emerging and Selected Topics in Industrial Electronics. He was Invited Speaker at the Gordon Research Conference on Advanced Health Informatics (AHI2018), General Chair of IEEE ES2017 and General Co-Chair of IEEE WFCS2021. He was awarded the "2016 Inventor of the Year Award" and “2018 Inventor of the Year Award” by ABB Corporate Research Sweden. 
\end{IEEEbiographynophoto}

\end{document}